\documentclass[
    ,final            
  ]
  {aipproc}

\layoutstyle{8x11single}
\renewcommand\XFMaddressmarkstyle{\alph}

\newcommand{\MJ}      {{\sc{Majo\-ra\-na}}}
\newcommand{\DEM}      {{\sc{Dem\-on\-stra\-tor}}}
\newcommand{\nonubb}  {0 \nu \beta \beta}
\newcommand{\twonubb}  {2 \nu \beta \beta}

\def\MJDEMbf{\bfseries{\scshape{Majorana Demonstrator}}}

\begin{document}

\title{The \MJDEMbf: A Search for Neutrinoless Double-beta Decay of Germanium-76}


\classification{23.40.-s, 14.60.Pq, 29.40.Wk}
\keywords      {neutrino, double-beta, germanium}

\newcommand{\alberta}{Centre for Particle Physics, University of Alberta, Edmonton, AB, Canada}
\newcommand{\blhill}{Department of Physics, Black Hills State University, Spearfish, SD, USA}
\newcommand{\ITEP}{Institute for Theoretical and Experimental Physics, Moscow, Russia}
\newcommand{\JINR}{Joint Institute for Nuclear Research, Dubna, Russia}
\newcommand{\lbnl}{Nuclear Science Division, Lawrence Berkeley National Laboratory, Berkeley, CA, USA}
\newcommand{\lanl}{Los Alamos National Laboratory, Los Alamos, NM, USA}
\newcommand{\uw}{Center for Experimental Nuclear Physics and Astrophysics, 
and Department of Physics, University of Washington, Seattle, WA, USA}
\newcommand{\unc}{Department of Physics and Astronomy, University of North Carolina, Chapel Hill, NC, USA}
\newcommand{\duke}{Department of Physics, Duke University, Durham, NC, USA}
\newcommand{\ncsu}{Department of Physics, North Carolina State University, Raleigh, NC, USA}
\newcommand{\ornl}{Oak Ridge National Laboratory, Oak Ridge, TN, USA}
\newcommand{\ou}{Research Center for Nuclear Physics and Department of Physics, Osaka University, Ibaraki, Osaka, Japan}
\newcommand{\pnnl}{Pacific Northwest National Laboratory, Richland, WA, USA}
\newcommand{\ttu}{Tennessee Tech University, Cookeville, TN, USA}
\newcommand{\sdsmt}{South Dakota School of Mines and Technology, Rapid City, SD, USA}
\newcommand{\sjtu}{Shanghai Jiaotong University, Shanghai, China}
\newcommand{\usc}{Department of Physics and Astronomy, University of South Carolina, Columbia, SC, USA}
\newcommand{\usd}{Department of Earth Science and Physics, University of South Dakota, Vermillion, SD, USA}
\newcommand{\ut}{Department of Physics and Astronomy, University of Tennessee, Knoxville, TN, USA}
\newcommand{\tunl}{Triangle Universities Nuclear Laboratory, Durham, NC, USA}


\author{S.R.~Elliott}{address={\lanl}}
\author{N.~Abgrall}{address={\lbnl}}
\author{E.~Aguayo}{address={\pnnl}}
\author{F.T.~Avignone~III}{address={\usc}, altaddress={\ornl}}
\author{A.S.~Barabash}{address={\ITEP}}
\author{F.E.~Bertrand}{address={\ornl}}
\author{M.~Boswell}{address={\lanl}} 
\author{V.~Brudanin}{address={\JINR}}
\author{M.~Busch}{address={\duke}, altaddress={\tunl}}
\author{A.S.~Caldwell}{address={\sdsmt}}
\author{Y-D.~Chan}{address={\lbnl}}
\author{C.D.~Christofferson}{address={\sdsmt}}
\author{D.C.~Combs}{address={\ncsu}, altaddress={\tunl}}
\author{J.A.~Detwiler}{address={\uw}}
\author{P.J.~Doe}{address={\uw}}
\author{Yu.~Efremenko}{address={\ut}}
\author{V.~Egorov}{address={\JINR}}
\author{H.~Ejiri}{address={\ou}}
\author{J.~Esterline}{address={\duke}, altaddress={\tunl}}
\author{J.E.~Fast}{address={\pnnl}}
\author{P.~Finnerty}{address={\unc}, altaddress={\tunl}}
\author{F.M.~Fraenkle}{address={\unc}, altaddress={\tunl}}
\author{A.~Galindo-Uribarri}{address={\ornl}}
\author{G.K.~Giovanetti}{address={\unc}, altaddress={\tunl}}
\author{J. Goett}{address={\lanl}}
\author{M.P.~Green}{address={\unc}, altaddress={\tunl}}
\author{J.~Gruszko}{address={\uw}}	
\author{V.E.~Guiseppe}{address={\usd}}
\author{K.~Gusev}{address={\JINR}}
\author{A.L.~Hallin}{address={\alberta}}
\author{R.~Hazama}{address={\ou}}
\author{A.~Hegai}{address={\lbnl}} 
\author{R.~Henning}{address={\unc}, altaddress={\tunl}}
\author{E.W.~Hoppe}{address={\pnnl}}
\author{S. Howard}{address={\sdsmt}} 
\author{M.A.~Howe}{address={\unc}, altaddress={\tunl}}
\author{K.J.~Keeter}{address={\blhill}}
\author{M.F.~Kidd}{address={\ttu}}
\author{O.~Kochetov}{address={\JINR}}
\author{S.I.~Konovalov}{address={\ITEP}}
\author{R.T.~Kouzes}{address={\pnnl}}
\author{B.D.~LaFerriere}{address={\pnnl}}
\author{J. Leon}{address={\uw}}	
\author{L.E.~Leviner}{address={\ncsu}, altaddress={\tunl}}
\author{J.C.~Loach}{address={\sjtu}}	
\author{S.~MacMullin}{address={\unc}, altaddress={\tunl}}
\author{R.D.~Martin}{address={\lbnl}}	
\author{S.~Mertens}{address={\lbnl}}
\author{L.~Mizouni}{address={\usc}, altaddress={\pnnl}}  
\author{M.~Nomachi}{address={\ou}}
\author{J.L.~Orrell}{address={\pnnl}}
\author{C. O'Shaughnessy}{address={\unc},altaddress={\tunl}} 
\author{N.R.~Overman}{address={\pnnl}}
\author{D.G.~Phillips II}{address={\ncsu}, altaddress={\tunl}}  
\author{A.W.P.~Poon}{address={\lbnl}}
\author{K.~Pushkin}{address={\usd}} 
\author{D.C.~Radford}{address={\ornl}}
\author{K.~Rielage}{address={\lanl}}
\author{R.G.H.~Robertson}{address={\uw}}
\author{M.C.~Ronquest}{address={\lanl}}	
\author{A.G.~Schubert}{address={\uw}}
\author{B. Shanks}{address={\unc}, altaddress={\tunl}}	
\author{T.~Shima}{address={\ou}}
\author{M.~Shirchenko}{address={\JINR}}
\author{K.J.~Snavely}{address={\unc}, altaddress={\tunl}}	
\author{N.~Snyder}{address={\usd}}	
\author{A.~Soin}{address={\pnnl}}	
\author{J.~Strain}{address={\unc}, altaddress={\tunl}}
\author{A.M.~Suriano}{address={\sdsmt}} 
\author{V.~Timkin}{address={\JINR}}
\author{W.~Tornow}{address={\duke}, altaddress={\tunl}}
\author{R.L.~Varner}{address={\ornl}}  
\author{S. Vasilyev}{address={\ut}}
\author{K.~Vetter}{address={\lbnl}, altaddress={Department of Nuclear Engineering, University of California, Berkeley, CA, USA}}
\author{K.~Vorren}{address={\unc}, altaddress={\tunl}} 
\author{B.R.~White}{address={\ornl}}	
\author{J.F.~Wilkerson$\mathrm{^{i,}}$}{address={\unc}, altaddress={\ornl}}    
\author{W.~Xu}{address={\lanl}}  
\author{E.~Yakushev}{address={\JINR}}
\author{A.R.~Young}{address={\ncsu}, altaddress={\tunl}}
\author{C.-H.~Yu}{address={\ornl}}
\author{V.~Yumatov}{address={\ITEP}}

\begin{abstract}
The  {\sc Majorana} collaboration is searching for neutrinoless double beta decay using $^{76}$Ge, which has  been shown to have a number of advantages in terms of sensitivities and backgrounds. The observation of neutrinoless double-beta decay would show that lepton number is violated and that neutrinos are Majorana particles and would simultaneously provide information on neutrino mass. Attaining sensitivities for neutrino masses in the inverted hierarchy region, $15 - 50$ meV,  will require large, tonne-scale detectors with extremely low backgrounds, at the level of $\sim$1~count/t-y
 or lower in the region of the signal.  The  {\sc Majorana} collaboration, with funding support from DOE Office of Nuclear Physics and NSF Particle Astrophysics, is constructing the {\sc Demonstrator}, an array consisting of 40~kg of p-type point-contact high-purity germanium (HPGe) detectors, of which $\sim$30~kg will be enriched to 87\% in $^{76}$Ge.  The {\sc Demonstrator} is being constructed in a clean room laboratory facility at the 4850' level (4300 m.w.e.)
of the
Sanford Underground Research Facility (SURF) in Lead, SD. It utilizes a compact graded shield approach with the inner portion consisting of ultra-clean Cu that is being electroformed and machined underground. The primary aim of the {\sc Demonstrator} is to show the feasibility of a future tonne-scale measurement in terms of backgrounds and scalability. 

\end{abstract}

\maketitle



\begin{figure}
\includegraphics[width=0.6\textwidth]{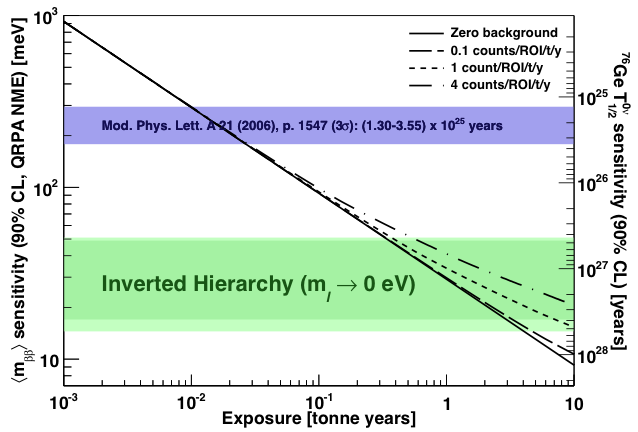} 
\caption{\label{fig:GeSens}90\% C.L. sensitivity as a function of exposure for $\nonubb$-decay searches in  $^{76}$Ge
under different background scenarios. The matrix element from Ref.~\citep{Sim09} was used to convert half-life to neutrino mass.  The upper shaded band shows the region where a signal would be detected should the Klapdor-Kleingrothaus claim \citep{kla06} be correct. $m_l$ in the lower shaded band refers to the lightest neutrino mass.
}
\end{figure}

\noindent
{\bf Scientific Motivation:~} Neutrinoless double-beta decay ($\nonubb$) is the most general, model independent method to search for lepton number violation and  correspondingly to determine the Dirac-Majorana nature of the neutrino\citep{Camilleri2008, avi08}.
Reaching the neutrino mass scale associated with the inverted mass hierarchy, $15 - 50$ meV, 
will require a half-life sensitivity on the order of 10$^{27}$~y,
corresponding to a signal of a few counts or less per tonne-year in the $\nonubb$ peak.
Observation of such a small signal will require  tonne-scale detectors with backgrounds in the region
of interest at or below  $\sim$1~count/t-y.
HPGe detectors have exceptionally good intrinsic energy resolution of $\sim$0.2\% at Q$_{\beta\beta}$ of 2039~keV, which for a 4 keV region of interest would correspond to a required background per keV of 
$2.5 \times 10^{-1}$ counts/(keV-t-y).
This excellent energy resolution ensures that background from the irreducible $\twonubb$  decay
 (T$_{1/2}$ = 1.8 x 10$^{21}$ y\citep{Agostini2013}) is negligible even for $\nonubb$ half lives beyond 10$^{27}$~y.
The sensitivity of a $\nonubb$ search increases with the exposure of the experiment, but ultimately depends on the achieved background level. This relationship is illustrated in Figure \ref{fig:GeSens}, where we have used the Feldman-Cousins definition of sensitivity in order to transition smoothly between the background-free and background-dominated regimes. Although this figure is drawn using experimental parameters  and theoretical nuclear matrix elements relevant for $\nonubb$ searches using $^{76}$Ge, the situation for other isotopes is not qualitatively different\citep{Robertson2013}. It may be concluded that achieving sensitivity to the entire parameter space for inverted-hierarchical  Majorana neutrinos would require,
using optimistic values of matrix elements and g$_A$, about
$\sim$5-10 tonne-years of exposure with a background rate of less than one count/t-y. Higher background levels would require significantly more mass to achieve the same sensitivity within a similar counting time.
We note that a convincing discovery that neutrinos are Majorana particles and that lepton number is violated will require the observation of  $\nonubb$ in multiple experiments using different $\nonubb$ isotopes.
\vspace {1.5 mm}

\noindent
{\MJDEMbf:}~ The \DEM\ exploits the established benefits of enriched high-purity germanium (HPGe) detectors, such as intrinsically low-background source material, understood enrichment chemistry, excellent energy resolution, and the ability to reject backgrounds using pulse-shape information. The primary technical challenge is the reduction of ionizing radiation backgrounds by about a factor 100 below what has been previously achieved. Specific goals of the \MJ\ \DEM\ are:

\vspace {-1mm}
\begin{itemize}
\item Demonstrate a path forward to achieving a background rate at or below 1 count/t-y in the 4-keV region of interest (ROI) around the  2039-keV  Q-value  for $^{76}$Ge $\nonubb$. This is required for tonne-scale germanium-based searches that will probe the inverted-hierarchy parameter space for $\nonubb$ decay.
\item Show technical and engineering scalability toward a tonne-scale instrument.
\item Test the Klapdor-Kleingrothaus et al. claim~\citep{kla06}. 
\item Perform searches for physics beyond the standard model, such as the search for dark matter and axions.
\end{itemize}
\vspace {-1mm}

To this end, the collaboration is building the \DEM, a modular instrument composed of two cryostats built from ultra-pure
electroformed copper,  each of which can house over 20~kg of  HPGe detectors 
contained in an ultra-low background structure that maximizes the
concentration of crystals while minimizing the amounts of structural materials\citep{Schubert2012, Wilkerson2012}.
The \DEM\ uses p-type point-contact (PPC) 
detectors that have masses in the range of 0.6-1.1~kg.
PPC style detectors were chosen after extensive R\&D by the collaboration for their advantages:  simple fabrication,
excellent pulse shape discrimination between $\nonubb$ events and backgrounds, and very low capacitance, providing a low-energy threshold that allows the reduction of background from cosmogenic $^{68}$Ge.
The array will contain about
 30~kg of detectors fabricated from 87\% enriched $^{76}$Ge and 10~kg of detectors fabricated from natural Ge.  This amount of enriched material is sufficient to achieve the physics and technical goals while optimizing cost,
and provides a systematic check of enriched vs. natural Ge.
Within the cryostats, the detectors are mounted in ``strings'', each consisting of 3-5 detectors.
A multi-stage Rn-suppressed glove box is used to build the strings and insert them into
the cryostats.
The modular approach allows us to assemble and optimize each cryostat independently, providing step-wise deployment with minimum interference on already operational detectors. 
The cryostats sit within a graded shield, where the inner passive shield will be constructed of electroformed and commercial high-purity copper, surrounded by high-purity lead, which itself is surrounded by an active muon veto and a layer of polyethylene to reduce the neutron flux.
The array is located in the Davis campus at the 4850 ft level of the Homestake Mine at SURF.

\begin{figure}
\includegraphics[width=0.6\textwidth]{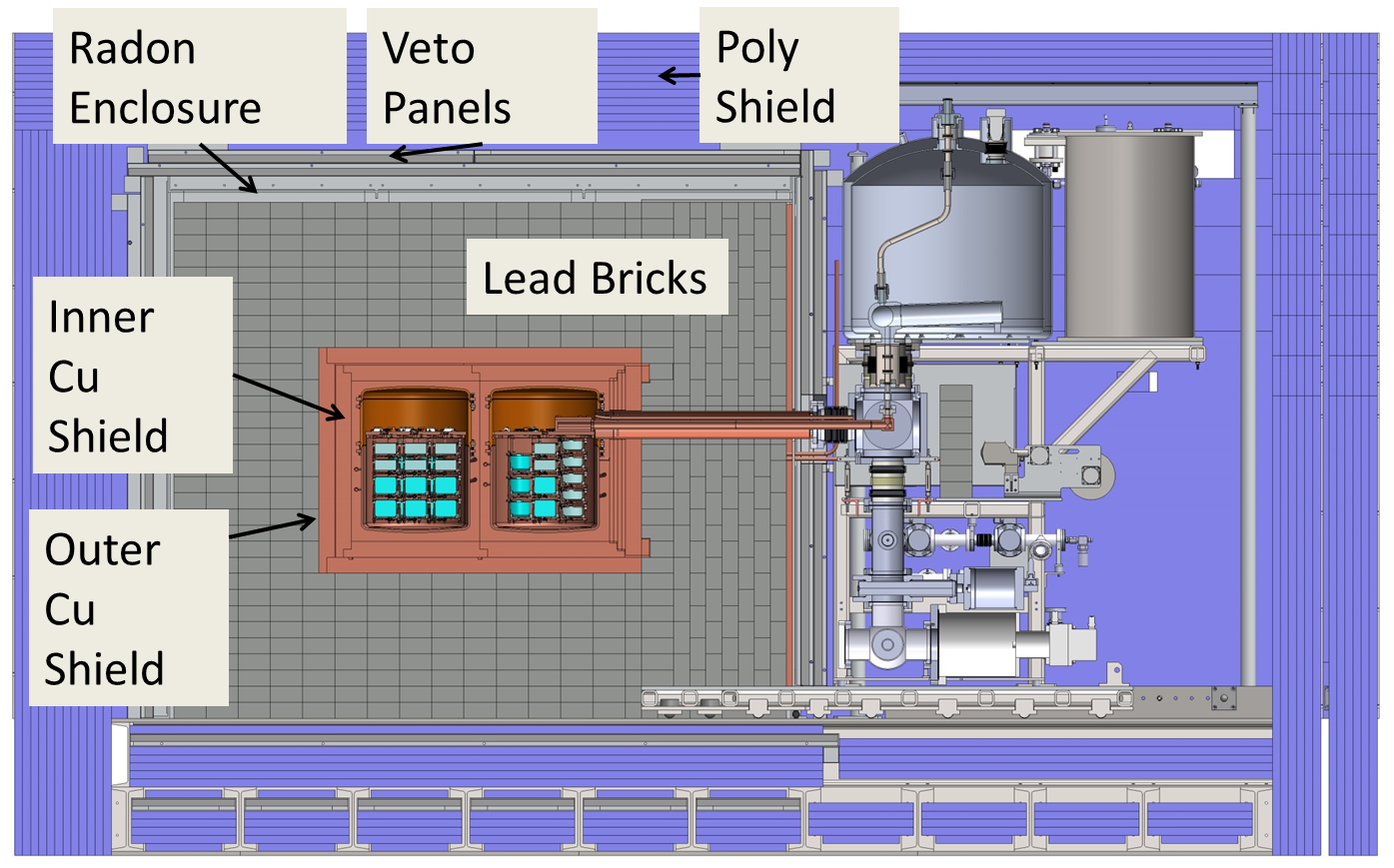} 
\caption{\label{fig:MJDCrossX}Cross-sectional view of the \MJ\ \DEM. Strings of detectors are visible in the two cryostats.
Cryostats are mounted on moveable transporters allowing independent assembly and testing before installation into the shield.  The cryostat vacuum and thermosyphon cooling systems are visible for the cryostat on the right. The Rn enclosure is a gas barrier whose internal volume will be continuously purged with liquid nitrogen boil-off gas to reduce Rn levels near the cryostats.
For scale, the inner Cu shield dimensions where the cryostats are inserted is 20" high and 30" in length.
}
\end{figure}

An essential aspect of the \DEM\ is the production and use of ultra-clean Cu. In typical materials uranium (U) and thorium (Th) decay-chain contaminants are found at levels of $\mu$g/g to ng/g, which will produce unacceptable backgrounds in the \DEM.
Electroforming copper in
a carefully-controlled and clean environment allows one to produce copper with U and Th below the level of $10^{-12}$ g/g\citep{Hoppe2009}.  The copper being produced by \MJ\ has about ten times lower U and Th impurities than commercial electroformed copper, with a projected activity of 0.3 $\mu$Bq/kg for Th or lower.  To avoid cosmogenic activation of the most sensitive parts, the copper is being produced at an underground (UG) production facility at SURF 
and at a shallow facility at Pacific Northwest National Laboratory, and is being machined UG in an ultra-clean machine shop installed and operated by the collaboration. Copper has mechanical, thermal, and electrical properties that are suitable for the \DEM's cryostats, detector mounts, and inner shield.

\vspace {1.5 mm}

\noindent
{\bf Status:~} Having obtained beneficial occupancy of its Davis campus UG laboratories in May 2012, the collaboration has completed outfitting the labs, established cleanliness (in the detector room better than class 500), and is proceeding with the construction and assembly of the array. The UG electroforming laboratories, which started operation in the summer of 2011, have now produced more than 75\% of the required copper.  The 42.5~kg of 87\% enriched $^{76}$Ge has been reduced from GeO$_2$ and refined to electronic grade Ge with a yield of 98\%. Ten enriched PPC detectors with a total mass of 9.5 kg have been produced by ORTEC, with nine now transported to SURF.  A prototype cryostat, the same as the ultra-clean cryostats but fabricated from commercial copper,  has been assembled and operated with its associated vacuum system. Two strings of natural Ge detectors have been built in the glove boxes and are undergoing testing.  Cryostat 1 has been machined from the ultra-clean electroformed Cu, and parts and components are being assembled.  Samples obtained from  all materials being used in the \DEM\ are being assayed. Slow control systems and their associated sensors are in continuous operation in both UG laboratories.  Data acquisition systems for detector acceptance testing, string testing, and the main array are operational.

The prototype cryostat will be commissioned this summer.  Cryostat 1, which will contain  strings of both enriched and natural Ge detectors, is scheduled to be commissioned in early 2014. Cryostat 2, which is expected to contain all enriched detectors, is scheduled to be be completed towards the end of 2014. The full array should be in operation in 2015.  The \DEM\ will be operated for about 3 years, collecting  $\sim$100 kg-years of exposure.  
 
The GERDA\citep{Agostini2013} collaboration is pursuing a similar program using a novel large liquid-argon shield. They are in the process of constructing Phase II of GERDA, which will consist of up to $\sim$35~kg of enriched detectors with a similar background goals as the  {\sc Majorana} {\sc Demonstrator}. If {\sc Majorana} and/or GERDA are able to demonstrate that the requisite backgrounds are achievable, then the collaborations plan to jointly pursue building a tonne-scale $^{76}$Ge array, selecting the best features and capabilities of the two current experiments. GERDA has recently announced a $\nonubb$ half-life limit of $3 x 10^{25}$ y~\citep{Agostini2013a}, which contradicts the previous claim~\citep{kla06}.


\begin{theacknowledgments}
We acknowledge support from the Office of Nuclear Physics in the DOE Office of Science under grant 
numbers DE-AC02-05CH11231, DE-FG02-97ER41041, DE-FG02-97ER41033, DE-FG02-97ER4104, 
DE-FG02-97ER41042, DE-SCOO05054, DE-FG02-10ER41715, and DE-FG02-97ER41020. We acknowledge support 
from the Particle and Nuclear Astrophysics Program of the National Science Foundation through grant 
numbers PHY-0919270, PHY-1003940, 0855314, PHY-1202950, MRI 0923142 and 1003399. We gratefully acknowledge support from the 
Russian Foundation for Basic Research, grant No. 12-02-12112. We gratefully acknowledge the support of the U.S. Department 
of Energy through the LANL/LDRD Program. 
\end{theacknowledgments}


\bibliographystyle{aipproc.bst}
\bibliography{DoubleBetaDecay.bbl}

\begin{thebibliography}{10}
\expandafter\ifx\csname natexlab\endcsname\relax\def\natexlab#1{#1}\fi
\providecommand{\enquote}[1]{``#1''}
\expandafter\ifx\csname url\endcsname\relax
  \def\url#1{\texttt{#1}}\fi
\expandafter\ifx\csname urlprefix\endcsname\relax\def\urlprefix{URL }\fi
\providecommand{\eprint}[2][]{\url{#2}}

\bibitem[Camilleri et~al.(2008)]{Camilleri2008}
L.~Camilleri, E.~Lisi, and J.~F. Wilkerson, \emph{Ann. Rev. Nucl. Part. Sci.}
  \textbf{58}, 343 (2008).

\bibitem[Avignone et~al.(2008)]{avi08}
F.~T.~{\protect III}. Avignone, S.~R. Elliott, and J.~Engel, \emph{Rev. Mod.
  Phys.} \textbf{80}, 481 (2008).

\bibitem[Agostini et~al.(2013{\natexlab{a}})]{Agostini2013}
M.~Agostini, et~al., \emph{J. Phys. G:Nucl. Part. Phys.} \textbf{40}, 035110
  (2013{\natexlab{a}}).

\bibitem[Robertson(2013)]{Robertson2013}
R.~G.~H. Robertson, \emph{Mod. Phys. Lett. A} \textbf{28}, 1350021 (2013).

\bibitem[{\protect \v{S}}imkovic et~al.(2009)]{Sim09}
F.~{\protect \v{S}}imkovic, A.~Faessler, H.~M{\protect\"{u}}ther, V.~Rodin, and
  M.~Stauf, \emph{Phys. Rev. C} \textbf{79}, 055501 (2009).

\bibitem[Klapdor-Kleingrothaus and Krivosheina(2006)]{kla06}
H.~V. Klapdor-Kleingrothaus, and I.~V. Krivosheina, \emph{Mod. Phys. Lett. A}
  \textbf{21}, 1547 (2006).

\bibitem[Schubert et~al.(2012)]{Schubert2012}
A.~G. Schubert, et~al., \emph{AIP Conf. Proc.} \textbf{1441}, 480 (2012),
  \eprint{arXiv:1109.1567}.

\bibitem[Wilkerson et~al.(2012)]{Wilkerson2012}
J.~F. Wilkerson, et~al., \emph{J. Phys. Conf. Ser.} \textbf{375}, 042010
  (2012), 12th International Conference on Topics in Astroparticle and
  Underground Physics, TAUP2011, presented by J.F. Wilkerson.

\bibitem[Hoppe et~al.(2009)]{Hoppe2009}
E.~W. Hoppe, et~al., \emph{J. Radioanal. Nucl. Chem.} \textbf{282}, 315 (2009).

\bibitem[Agostini et~al.(2013{\natexlab{b}})]{Agostini2013a}
M.~Agostini, et~al., Results on neutrinoless double beta decay of $^{76}$ge
  from {\sc gerda} phase i (2013{\natexlab{b}}), \eprint{arXiv:1307.4720}.

\end{thebibliography}

\end{document}